\documentclass[paper]{JHEP3} 
\usepackage{epsfig}
\usepackage{amsmath}

\vspace{2cm}
\def\lsi{\raise0.3ex\hbox{$<$\kern-0.75em\raise-1.1ex\hbox{$\sim$}}}
\def\gsi{\raise0.3ex\hbox{$>$\kern-0.75em\raise-1.1ex\hbox{$\sim$}}}

\newcommand{\Dsl}{{D\hspace{-7pt}{/}}}

\title{
The equation of state in lattice QCD: 
with physical quark masses towards the continuum limit}
\author{
Yasumichi Aoki$^a$, Zoltan Fodor$^{a,b}$, Sandor D.~Katz$^b$
and Kalman K.~Szab\'o$^a$\\
$^a$Department of Physics, University of Wuppertal, Gauss 20, D-42119, 
Germany\\
$^b$Institute for Theoretical Physics, E\"otv\"os University, P\'azm\'any
1, H-1117 Budapest, Hungary\\
}

\abstract{ The equation of state of QCD at vanishing chemical potential as a
function of temperature is determined for two sets of lattice spacings. Coarser
lattices with temporal extension of $N_t$=4 and finer lattices of $N_t$=6 are
used.  Symanzik improved gauge and stout-link improved staggered fermionic
actions are applied. The results are given for physical quark masses both for
the light quarks and for the strange quark.  Pressure, energy density, entropy
density, quark number susceptibilities and the speed of sound are presented.  
}

\preprint{\heplat{0510084}}

\keywords{ Thermal Field Theory,  Lattice QCD}

\begin{document}

\section{Introduction}

QCD at vanishing chemical potential and at increasing temperatures (T)
undergoes a transition at $T$=$T_c$. In the cold, hadronic phase the dominant
degrees of freedom are colourless hadrons, whereas in the high 
temperature, plasma phase the dominant degrees of freedom are
coloured objects (though the type of the transition --first order,
second order or crossover-- is not unambiguously determined yet, we use
the expression `phase' to characterize the system with different
dominant degrees of freedom). The equilibrium description as a function
of the temperature is given by the equation of state (EoS).
The complete determination of the EoS needs non-perturbative inputs,
out of which lattice simulations is the most 
systematic approach. In lattice simulations there are two obvious
difficulties, which are connected with 1.) the physical quark-mass limit 
and 2.) the continuum limit.
\begin{enumerate}
 \renewcommand{\labelenumi}{ad \theenumi.)}
 \item The masses of the up and down quarks 
($m_{ud}$) are small compared to the QCD scale. Since the computational 
costs increase sharply for small masses, most of the simulations are done 
at a larger mass than the physical one.
 \item The discretization of space-time through the lattice 
introduces the lattice spacing `$a$'. The continuum results are obtained
in the $a\rightarrow$0 limit. Again, the computational
costs increase sharply for small lattice spacings (for the EoS
it scales approximately as 1/$a^{13}$, which is a stronger growth of
CPU-costs, than for typical T=0 simulations e.g. spectroscopy). 
In order to approach the continuum limit fast, 
improved actions are used. They all give
the same QCD action in the continuum limit. However, for non-vanishing
lattice spacing they have much less discretization artefacts than
the most straightforward unimproved actions. In this paper we will use
a Symanzik improved gauge and stout link improved fermionic action.
\end{enumerate}
  
The EoS has been determined in the continuum limit for the pure gauge
theory (in this case --quenched simulations-- 
the simulations
are particularly easy as there is no fermionic degree of freedom.
The simulated systems are equivalent to those where all the fermions
are infinitely heavy, thus, is far from the physical situation)
\footnote{Note, however, that even in this relatively simple case 
there is still a few \% difference 
between the different approaches.}
\cite{Boyd:1996bx,Okamoto:1999hi,Namekawa:2001ih}.
Far less is known about the unquenched case (QCD with dynamical quarks).
There are published results for two-flavor QCD
using unimproved staggered \cite{Blum:1994zf,Bernard:1996cs}, 
and improved Wilson fermions \cite{AliKhan:2001ek}.
Few results are available for the 2+1 flavour case,
among which the study done by Karsch, Laermann and Peikert in the year 2000
\cite{Karsch:2000ps}, using p4-improved staggered fermions,
has often been used as the best result of EoS from lattice QCD.

In this letter we calculate the EoS of dynamical QCD and attempt
to improve on the result of Karsch, Laermann and 
Peikert by several means.\footnote{It is important to emphasize, that
these improvements are partly due to the increased CPU resources and
partly due to theoretical improvements over the last 5 years.}
\begin{enumerate}
 \renewcommand{\labelenumi}{\theenumi.)}
 \item
We use for the lightest hadronic degree of freedom the physical pion with
mass of $m_{\pi}\approx$140~MeV instead of having the unphysical
$m_{\pi}\approx$600~MeV of Karsch,Laermann and Peikert in Ref.
\cite{Karsch:2000ps}.
\item We use finer lattices than they \cite{Karsch:2000ps},
since our maximal temporal extension is $N_t$=6 instead of their temporal
extension of $N_t$=4.
\item As suggested for staggered QCD thermodynamics
\cite{Fodor:2002km,Csikor:2004ik}, 
we keep our system on the line of constant physics (LCP) instead
of increasing the physical quark mass with the temperature (if 
Karsch,Laermann and Peikert had cooled down e.g. two of their systems one at 
$T$=3$T_c$ and one at $T$=0.7$T_c$ down to T=0, the first system 
would have had approximately 4 times larger quark masses -two times larger pion masses- 
than the second one; this unphysical
choice is known to lead systematics, which are comparable to the 
difference between the interacting and non-interacting plasma).
\item For the staggered formulation of quarks the physically almost
degenerate pion triplet has an unphysical non-degeneracy
(so-called taste violation). This mass
splitting $\Delta m_{\pi}^2$ vanishes in the continuum
limit as $a\rightarrow$0. Due to our smaller lattice spacing and
particularly due to our stout-link improved \cite{Morningstar:2003gk} action the 
splitting $\Delta m_{\pi}^2$ is much
smaller than their splitting in Ref. \cite{Karsch:2000ps}.
\item In their Ref. \cite{Karsch:2000ps} they used the approximate
R-algorithm \cite{Gottlieb:1987mq} in their simulations. 
This algorithm has an intrinsic parameter, the stepsize 
which, similarly to the lattice spacing, has to be extrapolated to zero.
None of the previous staggered lattice thermodynamic studies carried out
this extrapolation. Using the R-algorithm without stepsize
extrapolation leads to uncontrolled systematic errors. Instead of using
the approximate R-algorithm this work uses the exact RHMC-algorithm
(rational hybrid Monte-Carlo) \cite{Clark:2002vz,Clark:2003na}.
\item They used \cite{Karsch:2000ps} the string tension to set the
physical scale. This quantity, strictly speaking, does not 
exist in dynamical QCD (the string breaks and two mesons are produced \cite{Bali:2005fu}). 
Instead of this somewhat problematic
quantity we use the quark-antiquark potential
at intermediate distances. 
(This part of the potentials is theoretically sound \cite{Sommer:1994ce} and
it has been studied in the continuum limit 
\cite{Aubin:2004wf}).
\end{enumerate}

For completeness --and for fairness-- one should also mention that the 
analysis of \cite{Karsch:2000ps} has also advantages (though as
it will be shown, these advantages are also reached by the present
letter in some way or another).
\begin{enumerate}
 \renewcommand{\labelenumi}{\theenumi.)}
\item At infinitely large T both the action of \cite{Karsch:2000ps} and 
that of the present
work approaches the continuum result as 1/$N_t^2$, though the prefactor
for the action used by \cite{Karsch:2000ps} is smaller. 
Note,    
however, that an extrapolation based on our $N_t$=4 and 6 temporal extensions
would give even in this limit  roughly the same deviation from the 
continuum value as the action of Karsch, Laermann and Peikert at their
$N_t$=4 temporal extension. 
\item The thermodynamic limit is usually approached by large volumes V,
which can be characterized by the aspect ratio $r$=$N_s$/$N_t$ ($N_s$ is 
the spatial extension of the lattice). The aspect ratio of 
Ref. \cite{Karsch:2000ps} is 4. This is larger than $r$
of the present analysis, which was in most of the cases 3. 
Note; however, that at several points we used aspect ratios of
3 and 4 in order to check, that the uncertainties due to $r$=3
are smaller than our statistical errors.
\end{enumerate}

In addition there is --at least-- one uncertainty that neither
\cite{Karsch:2000ps} nor this work can really address. When
determining the EoS at fixed temporal extension (in \cite{Karsch:2000ps}
with $N_t$=4 and in this letter with $N_t$=4,6) the lattice spacing
changes when we change the temperature. Thus, the finite lattice
spacing effects are different at low and at high temperatures.
These uncertainties can be only resolved by continuum extrapolation.
Though we have the EoS on two different sets of lattice spacings and
one might attempt to do this extrapolation, it is fair to say that another
set of lattice spacings is needed ($N_t$=8). One of the reasons is, that
in the hadronic phase, where the integration for the pressure starts,
the lattice spacing is larger than 0.3 fm. In this region
the lattice artefacts can not be really controlled (and in this deeply
hadronic case it does not really help that an action is very 
good at asymptotically high temperatures, in the free non-interacting
gas limit \cite{Heller:1999xz}, as the p4 action of \cite{Karsch:2000ps} is).

There are other ongoing thermodynamics projects (c.f. 
\cite{Bernard:2005mf,Jung:2005yb}) 
improving on previous results. 
Both the MILC and the RBC-Bielefeld collaborations presented new results
at the lattice conference. The MILC collaboration studies the equation of
state along LCP's with two light quark masses (0.1 and 0.2 times $m_s$) at
$N_t$=4 and 6 lattices using asqtad improved staggered fermionic action 
\cite{Bernard:2005mf}. The RBC-Bielefeld Collaboration developed the p4fat7 
action to
reduce taste symmetry violation. They study the $N_f=3$ phase transition
with several quark masses down to a pion mass of $\approx$ 320 MeV on
$N_t$=4 and 6 lattices \cite{Jung:2005yb}. Both works use the R algorithm.

The paper is organized as follows. In Section~2 
our definition for the Symanzik improved gauge and for the stout-link 
improved fermionic action is presented. We show our result on 
the line of constant physics and discuss how it was obtained.
We demonstrate that our choice of action has 
small
taste violation, when compared to other actions. The advantages
of our exact RHMC simulation algorithm are discussed, too. Some
details on the simulation points are summarized. 
Section~3 shortly discusses
the methods (c.f. \cite{Engels:1990vr}) to determine the EoS and presents
results at our two lattice spacings for the pressure, energy density, 
entropy density, quark
number susceptibilities and the speed of sound.
In Section~4 we summarize and conclude.

\section{Lattice action, simulations and the line of constant physics}

This section contains several technical details. Readers, who
are not interested in these details, can simply skip to the next section.
First we give our definition for the Symanzik improved gauge 
and for the stout-link improved fermionic action. 
We demonstrate that our choice of stout-link improved staggered fermionic
action has  
small
taste violation, when compared to other staggered actions used in the literature to determine the EoS of QCD. The advantages
of our exact RHMC simulation algorithm are emphasized. 
We discuss the importance of the LCP
and show how to determine it by simulating in the three-flavour
theory and by using the pseudoscalar and vector meson masses.
Some details on the simulation points are summarized. 
\FIGURE[t]{
\hspace{2cm}
\epsfig{file=del4_cmp3.eps,width=7.2cm}
\hspace{2cm}
\caption{
 Pion mass splitting $\Delta_\pi=(m_\pi'^2-m_\pi^2)/T_c^2$ as a 
 function of $(m_\pi/T_c)^2$. The lattice spacings are the same as those at the
 finite temperature transition point. 
The mass of the Goldstone pion is denoted by $m_\pi$, that of the first
 non-Goldstone mode is by $m_\pi'$. The horizontal blue line corresponds to the physical value of
$(m_\pi/T_c)^2$, where $T_c= 173$MeV was assumed \cite{Karsch:2000kv}. 
The taste violation of our stout-link improved action
is much smaller than that of the unimproved action and even somewhat smaller than 
that of the asqtad action at the same $N_t$. 
}
\label{taste}
}

Isotropic lattice couplings are used, thus the lattice spacings are identical
in all directions.  The lattice action we used has the following form:
\begin{eqnarray}
\label{action} S & = & S_g + S_f,\\ S_g & = & \sum_x
\frac{\beta}{3} (  c_0  \sum_{\mu>\nu} W_{\mu,\nu}^{1\times 1}(x) + c_1
\sum_{\mu\ne\nu} W_{\mu,\nu}^{1\times 2}(x) ),\\ S_f & = & \sum_{x,y} \{
\overline{\eta}_{ud}(x) [\Dsl(U^{stout})_{xy}+m_{ud}\delta_{x,y}]^{-1/2}
\eta_{ud}(y) \nonumber\\ &  & \mbox{\hspace{2pt}} +\mbox{\hspace{2pt}}
\overline{\eta}_{s}(x) [\Dsl(U^{stout})_{xy}+m_{s}\delta_{x,y}]^{-1/4}
\eta_{s}(y)\}, 
\end{eqnarray} 
where $W_{\mu,\nu}^{1\times 1}$,
$W_{\mu,\nu}^{1\times 2}$ are real parts of the traces of the ordered products
of link matrices along the $1\times 1$, $1\times 2$ rectangles in the $\mu$,
$\nu$ plane.  The coefficients satisfy $c_0+8c_1=1$ and $c_1=-1/12$ for the
tree-level Symanzik improved action.  $\eta_{ud}$ and $\eta_s$ are the
pseudofermion fields for $u$, $d$ and $s$ quarks.  $\Dsl(U^{stout})$ is the
four-flavor 
staggered Dirac matrix with stout-link improvement \cite{Morningstar:2003gk}.
Let us also note here, that we use the 4th root trick in eq. (\ref{action}),
which might
lead to problems of locality.

Our staggered action at a given $N_t$ yields the same limit for the pressure at infinite
temperatures as the standard unimproved action. There are various techniques
improving the high temperature scaling. However, usually an extrapolation based
on $N_t$ and $N_t+2$ with standard staggered action gives a better high T behavior
for the pressure than p4 \cite{Heller:1999xz} or asqtad
\cite{Lepage:1998vj,Orginos:1999cr} action with $N_t$. 
Since our choice of action is about an order of magnitude faster than
e.g. p4, we decided to use this less impoved action, with which our CPU
resources made it possible to study two lattice spacings ($N_t$=4 and 6).

Staggered fermions have an unconvenient property: they violate taste symmetry
at finite lattice spacing. Among other things this violation results in a
splitting in the pion spectrum, which should vanish in the continuum limit.
The stout-link improvement makes the staggered fermion taste symmetry
violation small already at moderate lattice spacings.  We found that a
stout-smearing level of $N_{smr}$=2 and smearing parameter of $\rho$=0.15 are
the optimal values of the smearing procedure.  In order to illustrate the
advantage of the stout-link action Figure \ref{taste} compares the taste
violation in different approaches of the literature, which were used to
determine the EoS of QCD.  Results on the pion mass splitting for unimproved
(used by Ref.~\cite{Blum:1994zf,Bernard:1996cs})
\footnote{We performed simulations to obtain the MILC standard action value at $m_q=0.0125$, $\beta_c=5.415$.}, 
p4 improved (used by
Ref.~\cite{Karsch:2000ps,Peikert:2000:thesis}), 
asqtad improved (used by \cite{Bernard:2005mf,MILCpriv})
and stout-link improved (this
work) staggered fermions are shown. The parameters were chosen to be the ones
used by the different collaborations at the finite temperature transition
point.

In previous staggered analyses the gauge configurations were produced by the
R-algorithm \cite{Gottlieb:1987mq} at a given stepsize. These studies were
carried out usually at one stepsize, which is 1/2 or 2/3 of the light quark
mass.  The stepsize is an intrinsic parameter of the algorithm, which has to
be extrapolated to zero.  None of the previous staggered lattice
thermodynamics studies performed this extrapolation. Using the R-algorithm
without stepsize extrapolation leads to uncontrolled systematic errors.  E.g.
let us look at the difference (on $N_t$=6 lattices at intermediate $\beta$)
between the extrapolated plaquette value and the value obtained at stepsize
which is 2/3 of the light quark mass. This difference is larger than the total
contribution of the plaquette to the pressure. Clearly, such a technique can
not be used.

\FIGURE[t]{
\hspace{2cm}
\epsfig{file=m_s-beta.eps,width=7.2cm}
\hspace{2cm}
\caption{\label{LCP}
The line of constant physics. The result was obtained by using the
$\phi$ and K masses (see text). The strange quark mass in lattice
units is shown as a function of $\beta$. In the rest of our analysis we 
use light quark masses
of $m_{ud}$=$m_s$/25.
}
}

Instead of using the approximate R-algorithm this work uses the exact
RHMC-algorithm (rational hybrid Monte-Carlo) \cite{Clark:2002vz,Clark:2003na}.
This technique approximates the fractional powers of the Dirac operator by
rational functions.  Since the condition number of the Dirac operator changes
as we change the mass, one should determine the optimal rational approximation
for each quark mass. Note however, that this should be done only once, and the
obtained parameters of these functions can be used in the entire configuration
production. Our choices for the rational approximation were as good as few
times the machine precision for the whole range of the eigenvalues of the
Dirac operator. 
 
Let us discuss the determination of the LCP.  The LCP is defined as
relationships between the bare lattice parameters ($\beta$ and lattice bare
quark masses $m_{ud}$ and $m_s$). These relationships express that the physics
(e.g. mass ratios) remains constant, while changing any of the parameters.
It is important to emphasize that the LCP is unambiguous (independent of the
physical quantities, which are used to define the above relationships) only in
the continuum limit ($\beta\rightarrow\infty$).  For our lattice spacings
fixing some relationships to their physical values means that some other
relationships will slightly deviate from the physical one.  In thermodynamics
the relevance of LCP comes into play when the temperature is changed by
$\beta$ parameter. Then adjusting the mass parameters ($m_{ud}$ and $m_s$) is
an important issue, neglecting this in simulations can lead to several \%
error in the EoS.

A particularly efficient (however only approximate, see later) way to obtain an LCP is by using simulations with 
three degenerate flavors with lattice quark mass $m_q$.
The leading order chiral perturbation theory implies 
the mass relation for $s\bar{s}$ mesons.
The strange quark mass is tuned accordingly, as
\begin{equation}
 m_{PS}^2/m_{V}^2|_{m_q=m_s} = (2m_K^2-m_\pi^2)/m_\phi^2,
  \label{eq:tl chpt}
\end{equation}
where $m_{PS}$ and $m_{V}$ are the pseudoscalar and vector meson 
masses in the simulations with three degenerate quarks.
The light quark mass is calculated using the ratio $m_{ud}=m_s/25$
obtained by experimental mass input in the chiral perturbation theory
\cite{Gasser:1984gg}.
We obtain $m_s(\beta)$ as shown in Figure \ref{LCP}.

Our approach using eq. (\ref{eq:tl chpt}) is appropriate if in the $n_f$=2+1 
theory the vector meson mass depends only weakly on the light quark masses
and the chiral perturbation theory for meson masses works upto the strange
quark mass.
After applying the LCP we cross-checked the obtained spectrum of the 
$n_f$=2+1 simulations. 
The resulting pseudoscalar (pion, kaon) and phi mass ratios agree 
with the experimental values on the 5--10\% level.
In dynamical simulations approximately 
10\% effect on
$r_0$ is observed by changing the light quark masses 
from the strange mass to the physical
limit. Thus, this slight mismatch of the spectrum results in a
subdominant error on our overall scale, which is less or around 
2\%.

The determination of the EoS needs quite a few simulation points. Results
are needed on finite temperature lattices ($N_t$=4 or 6) and on zero
temperature lattices ($N_t \gg$ 4 or 6) at several $\beta$ values (we
used 16 different $\beta$ values for $N_t$=4 and 14 values for $N_t$=6). Since
our goal is to determine the EoS for physical quark masses we have to
determine quantities in this small physical quark mass limit
(we call these $\beta$ dependent bare light quark masses $m_{ud}(phys)$). 
\TABLE{\label{points}
\begin{tabular}{|c|c|c|c|c|c||c|c|c|c|c|c|}
\hline
$\beta$    & $m_s$ & T=0  & \# & T$\neq$0& \# &
$\beta$    & $m_s$ & T=0  & \# & T$\neq$0& \#  \\ 
\hline 
3.000 & 0.1938 & 16$^3$$\cdot$16& 4  &12$^3$$\cdot$4& 9 &3.450 & 0.1507 & 16$^3$$\cdot$32&	29  & 18$^3$$\cdot$6 & 120	\\
3.150 & 0.1848 & 16$^3$$\cdot$16& 4  &12$^3$$\cdot$4& 9 &3.500 & 0.1396 & 16$^3$$\cdot$32&	33  & 18$^3$$\cdot$6 & 156	\\
3.250 & 0.1768 & 16$^3$$\cdot$16& 4  &12$^3$$\cdot$4& 9 &3.550 & 0.1235 & 16$^3$$\cdot$32&	30  & 18$^3$$\cdot$6 & 133	\\
3.275 & 0.1742 & 16$^3$$\cdot$16& 4  &12$^3$$\cdot$4& 9 &3.575 & 0.1144 & 16$^3$$\cdot$32&	28  & 18$^3$$\cdot$6 & 151	\\
3.300 & 0.1713 & 16$^3$$\cdot$16& 4  &12$^3$$\cdot$4& 9 &3.600 & 0.1055 & 16$^3$$\cdot$32&	31  & 18$^3$$\cdot$6 & 158	\\
3.325 & 0.1683 & 16$^3$$\cdot$16& 4  &12$^3$$\cdot$4& 9 &3.625 & 0.0972 & 16$^3$$\cdot$32&	33  & 18$^3$$\cdot$6 & 144	\\
3.350 & 0.1651 & 16$^3$$\cdot$16& 4  &12$^3$$\cdot$4& 9 &3.650 & 0.0895 & 16$^3$$\cdot$32&	30  & 18$^3$$\cdot$6 & 160	\\
3.400 & 0.1583 & 16$^3$$\cdot$16& 3  &12$^3$$\cdot$4& 9 &3.675 & 0.0827 & 16$^3$$\cdot$32&	32  & 18$^3$$\cdot$6 & 178 	\\
3.450 & 0.1507 & 16$^3$$\cdot$32& 29  &12$^3$$\cdot$4& 9 &3.700 & 0.0766 & 16$^3$$\cdot$32&	33  & 18$^3$$\cdot$6 & 174	\\
3.500 & 0.1396 & 16$^3$$\cdot$32& 33  &12$^3$$\cdot$4& 9 &3.750 & 0.0666 & 16$^3$$\cdot$32&	35  & 18$^3$$\cdot$6 & 140	\\
3.550 & 0.1235 & 16$^3$$\cdot$32& 30  &12$^3$$\cdot$4& 9 &3.800 & 0.0589 & 20$^3$$\cdot$40&	26  & 18$^3$$\cdot$6 & 158	\\
3.600 & 0.1055 & 16$^3$$\cdot$32& 31  &12$^3$$\cdot$4& 9 &3.850 & 0.0525 & 20$^3$$\cdot$40&	23  & 18$^3$$\cdot$6 & 157	\\
3.650 & 0.0895 & 16$^3$$\cdot$32& 30  &12$^3$$\cdot$4& 9 &3.930 & 0.0446 & 24$^3$$\cdot$48&	6   & 18$^3$$\cdot$6 & 171	\\
3.700 & 0.0766 & 16$^3$$\cdot$32& 33  &12$^3$$\cdot$4& 9 &4.000 & 0.0401 & 28$^3$$\cdot$56&	4   & 18$^3$$\cdot$6 & 166	\\
3.850 & 0.0525 & 20$^3$$\cdot$40& 23  &12$^3$$\cdot$4& 9 & & & & & & \\
4.000 & 0.0401 & 28$^3$$\cdot$56& 4   &12$^3$$\cdot$4& 9 & & & & & & \\
\hline                                     
\end{tabular}                              
\caption{                                  
Summary of our simulation points. For the physical light quark masses 
(we call them $m_{ud}(phys)$) 
25 times smaller values were taken than for the strange mass. T$\neq$0
simulations were performed with the above $m_s$ and $\beta$ pairs,
and at 5 different $m_{ud}$ values: \{1,3,5,7,9\}$\cdot m_{ud}(phys)$.
T=0 simulations were performed with the above $m_s$ and $\beta$ pairs,
but at 4 different $m_{ud}$ values: \{3,5,7,9\}$\cdot m_{ud}(phys)$.
The total number of trajectories divided by 100 are collected
in the \# columns. 
The left column shows the $N_t$=4, whereas the right column shows
the $N_t$=6 data. (For an explanation of our labeling see the text.)
}
}

For our finite temperature simulations ($N_t$=4,6) we used physical quark
masses.  The spatial sizes were always at least 3 times the temporal sizes.
For the whole $\beta$ range on $N_t=4$ we checked that by increasing the
$N_s/N_t$ ratio from 3 to 4 the results remained the same within our
statistical uncertainties. 

In the chirally broken phase (our zero temperature simulations, thus lattices
for which $N_t \gg$ 4 or 6, belong always to this class) chiral perturbation
theory can be used to extrapolate by a controlled manner to the physical light
quark masses.  Therefore 
for most of our simulation points\footnote{In the $\beta=3.0..3.4$ range the
$T=0$ simulations were carried out at $m_{ud}(phys)$.}
we used four pion masses ($m_\pi\approx$235, 300, 355
and 405 MeV), which were somewhat larger than the physical one. (To simplify
our notation in the rest of this paper we label these points as 3,5,7 and 9
times $m_{ud}(phys)$.) It turns out that the chiral condensates at all the
four points can be fitted by linear function of pion mass squared with good
$\chi^2$. (Later we will show, that only the chiral condensate is to be
extrapolated to get the EoS at the physical quark mass.) The volumes were
chosen in a way, that for three out of these four quark masses the spatial
extentions of the lattices were approximately equal  or larger than four times
the correlation lengths of the pion channel.  We checked for a few $\beta$
values that increasing the spatial and/or temporal extensions of the lattices
results in the same expectation values within our statistical uncertainties.
(For 3$\cdot m_{ud}(phys)$ values the spatial lengths of the lattices were
only three times the correlation length of the pion channel.  However,
excluding this point from the extrapolations, the results do not change.) 

A detailed list of our
simulation points at zero and at non-zero temperature lattices
are summarized in Table \ref{points}. 

\section{Equation of state}

In this section
results for two sets of lattice spacings ($N_t$=4,6)
for the pressure, energy density, entropy density, quark
number susceptibilities and for the speed of sound are presented.

We shortly review the integral technique to obtain the pressure
\cite{Engels:1990vr}. For large homogeneous systems the pressure
is proportional to the logarithm of the partition
function: 
\begin{align}
\label{eq:pa}
pa^4=\frac{Ta}{V/a^3}\log Z(T,V)=\frac{1}{N_tN_s^3}\log Z(N_s,N_t;\beta,m_q).
\end{align}
(Index `q' refers to the ${ud}$ and $s$ flavors.)
The volume and temperature are connected to the spatial and temporal
extensions of the lattice:
\begin{align}
V=(N_sa)^3, && T=\frac{1}{N_ta}.
\end{align}  
The divergent zero-point energy has to be 
removed by subtracting the zero 
temperature ($N_t\to \infty$) part of 
eq. (\ref{eq:pa}). In practice the zero 
temperature subtraction 
is performed by using lattices with finite, but large $N_{t}$ (called $N_{t0}$, see Table \ref{points}). So the normalized 
pressure
becomes:
\begin{align}
\frac{p}{T^4}=N_t^4\left[ \frac{1}{N_tN_s^3}\log
 Z(N_s,N_t;\beta,m_q) - \frac{1}{N_{t0}N_{s0}^3}\log 
Z(N_{s0},N_{t0};\beta,m_q) \right].
\end{align}
With usual Monte-Carlo techniques one cannot measure $\log Z$ directly, but only its
derivatives with respect to the bare parameters of the lattice action.
Having determined the partial derivatives one
integrates in the
multi-dimensional parameter space:
\begin{equation}\label{integral}
\frac{p}{T^4}=
N_t^4\int^{(\beta,m_q)}_{(\beta_0,m_{q0})}
d (\beta,m_q)\left[
\frac{1}{N_tN_s^3}
\left(\begin{array}{c}
{\partial \log Z}/{\partial \beta} \\
{\partial \log Z}/{\partial m_{q}}
\end{array} \right )-
\frac{1}{N_{t0}N_{s0}^3}
\left(\begin{array}{c}
{\partial \log Z_0}/{\partial \beta} \\
{\partial \log Z_0}/{\partial m_q}
\end{array} \right )
\right],
\end{equation}
where $Z/Z_0$ are shorthand notations for $Z(N_s,N_t)/Z(N_{s0},N_{t0})$.
Since the integrand is a gradient, the result is by
definition independent of
the integration path. We need the pressure
along the LCP, thus it is convenient to measure the derivatives of $\log Z$
along the LCP and perform the integration over
this line in the $\beta$, $m_{ud}$ and $m_s$ parameter space. 
The lower limits of the integrations (indicated by $\beta_0$ and $m_{q0}$)
were set sufficiently below the transition point. By this choice
the pressure gets
independent of the starting point (in other words it vanishes
at small temperatures).  In the case of $2+1$ flavor staggered QCD
the derivatives of $\log Z$ with
respect to $\beta$ and $m_q$ are proportional to 
the expectation value of the gauge action ($\langle S_g \rangle$ c.f. eq. (\ref{action})) and to the 
chiral condensates ($\langle \bar{\psi}\psi_q \rangle $), respectively. 
Eq. (\ref{integral}) can be rewritten appropriately and the pressure
is given by (in this formula we write out explicitely the flavours):
\begin{equation}\label{pmu0}
\frac{p}{T^4}=
N_t^4\int^{(\beta,m_{ud},m_s)}_{(\beta_0,m_{ud0},m_{s0})}
d (\beta,m_{ud},m_s)\left[
\frac{1}{N_tN_s^3}
\left(\begin{array}{c}
\langle{\rm -S_g/\beta}\rangle \\
\langle\bar{\psi}\psi_{ud}\rangle \\
\langle\bar{\psi}\psi_{s}\rangle
\end{array} \right) 
-
\frac{1}{N_{t0}N_{s0}^3}
\left(\begin{array}{c}
\langle{\rm -S_g/\beta}\rangle_0 \\
\langle\bar{\psi}\psi_{ud}\rangle_0 \\
\langle\bar{\psi}\psi_{s}\rangle_0
\end{array} \right)\right], 
\end{equation}
where $\langle \dots \rangle _0$ means averaging on a $N_{s0}^3\cdot N_{t0}$ lattice.

The integral method was originally introduced for the pure gauge case, for
which the integral is one dimensional, it is performed along the $\beta$ axis.
Previous studies for staggered dynamical QCD (e.g.
\cite{Bernard:1996cs,Engels:1996ag,Karsch:2000ps}) used a one-dimensional
parameter space instead of performing it along the LCP. Note, that for full
QCD the integration should be performed along a LCP path in a
multi-dimensional parameter space. 
\TABLE{ \label{ta:c_values}
\hspace{2cm}
\begin{tabular}{|c||c|c|c|}
\hline
$N_t$   & p/$T^4$   & $c_s^2$   & $\chi/T^2$     \\
\hline
4       & 9.12    & 1/3   &  2.24 \\
6       & 7.86   &  1/3   &  1.86 \\
$\infty$   & 5.21   & 1/3  & 1 \\
\hline
\end{tabular}
\hspace{2cm}
\caption{
Summary of the results for the 2+1 flavor pressure, 
speed of sound and 1 flavor quark number susceptibility in the non-interacting
Stefan-Boltzmann limit. $\epsilon/T^4$ is 3 times, whereas $s/T^3$ is
4 times the normalized value of the pressure ($p/T^4$) in the Stefan-Boltzmann limit.
The first two lines gives the results
for $N_t$=4,6 and the third line contains the results 
in the continuum
(in the thermodynamic limit). 
}
}

Using appropriate thermodynamical relations one can obtain any thermal properties
of the system. For example the energy density ($\epsilon$), entropy density ($s$)
and
speed of sound ($c_s^2$)
can be derived as
\begin{align}
\label{eq:escs}
\epsilon = T(\partial p/\partial T)-p, && s = (\epsilon + p) T, && c_s^2
=\frac{dp}{d\epsilon}.
\end{align}
To be able to do theses derivatives one has to know the temperature along the LCP.
Since the temperature is connected to the lattice spacing as
\(
T=(N_t a)^{-1
},
\)
we need a reliable estimate on $a$.
The lattice spacings at different points of the LCP are
determined by first matching the static potentials for different $\beta$ values at an intermediate distance
for $m_{ud}=\{3,5\}m_{ud}(phys)$ quark masses, then extrapolating
the results to the physical quark mass.
Relating these distances to physical observables
(determining the overall scale in physical units) will be the topic of
a subsequent publication. 
We show the results as a function of $T/T_c$. The transition
temperature ($T_c$) is defined by the inflection point of the
isospin number 
susceptibility ($\chi_{I}$, see later).

To get the energy density the literature usually uses another quantity, namely
$\epsilon$-3$p$, which can be also directly measured on the lattice.  In our
analysis it turned out to be more appropriate to calculate first the pressure
directly from the raw lattice data (eq. (\ref{pmu0})) and then determine the
energy density and other quantities from the pressure (eq. (\ref{eq:escs})).
The reasons for that can be summarized as follows. As we discussed we perform
T$\neq$0 simulations with physical quark masses, whereas the subtraction terms
from T=0 simulations are extrapolated from larger quark masses. This sort of
extrapolation is adequate for the chiral condensates, for which chiral
perturbation techniques work well. Thus, one can choose an integration path
for the T=0 part of the pressure, which moves along a LCP at some larger
$m_{ud}$ (e.g. 9 times $m_{ud}(phys)$) and then at fixed $\beta$ goes down to
the physical quark mass. No comparable analogous technique is available for
the combination $\epsilon$-3$p$. 

We have also calculated the pressure for the larger quark masses. Plotting it
as a function of the temperature the differences between them are significant.
As a function of $T/T_c$ these differences are smaller, but still remain
statistically significant in the $1.2...2.0 T_c$ region.  
Note that statements on the mass dependence are only qualitative
since such an analysis requires the careful matching of the scales at
different quark masses. It is non-trivial and will be the topic of a
forthcoming paper.

\FIGURE[t]{
\epsfig{file=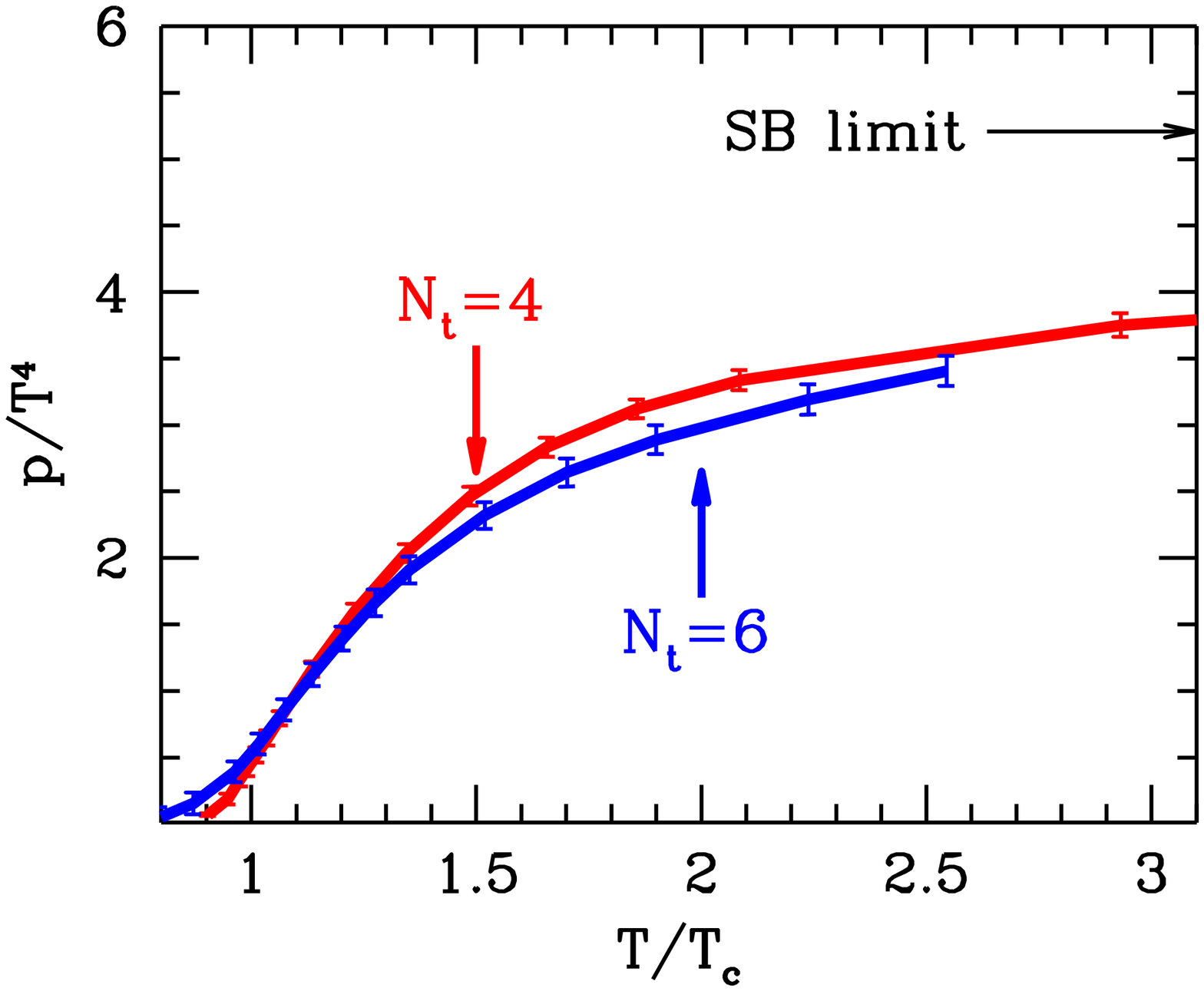,width=7.2cm}
\epsfig{file=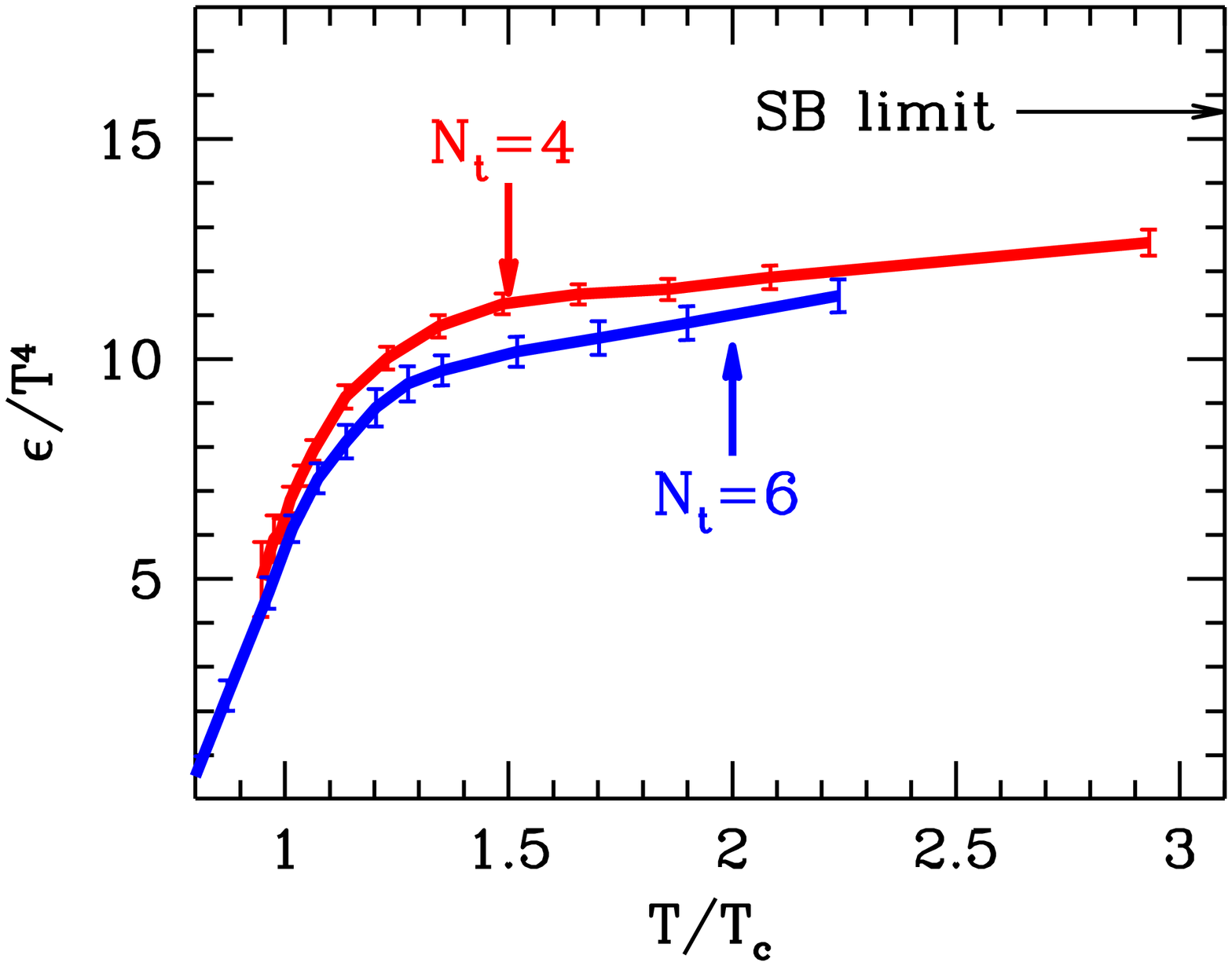,width=7.2cm}
\caption{ \label{eos_pe}
a.) The left panel shows the pressure $p$,
as a function of the temperature. Both $N_t$=4 (red, upper curve)
and $N_t$=6 (blue, lower curve) data are obtained along the LCP. They 
are normalized by $T^4$ and scaled by $c_{cont}/c_{N_t}$ (see text
and Table \ref{ta:c_values}). 
In order to lead the eye lines connect the data points.
b.) The right panel is the energy density ($\epsilon$), red (upper) and blue (lower) for $N_t$=4 and
6 respectively. This result was obtained directly from the pressure. 
}
}
\FIGURE{
\hspace{2cm}
\epsfig{file=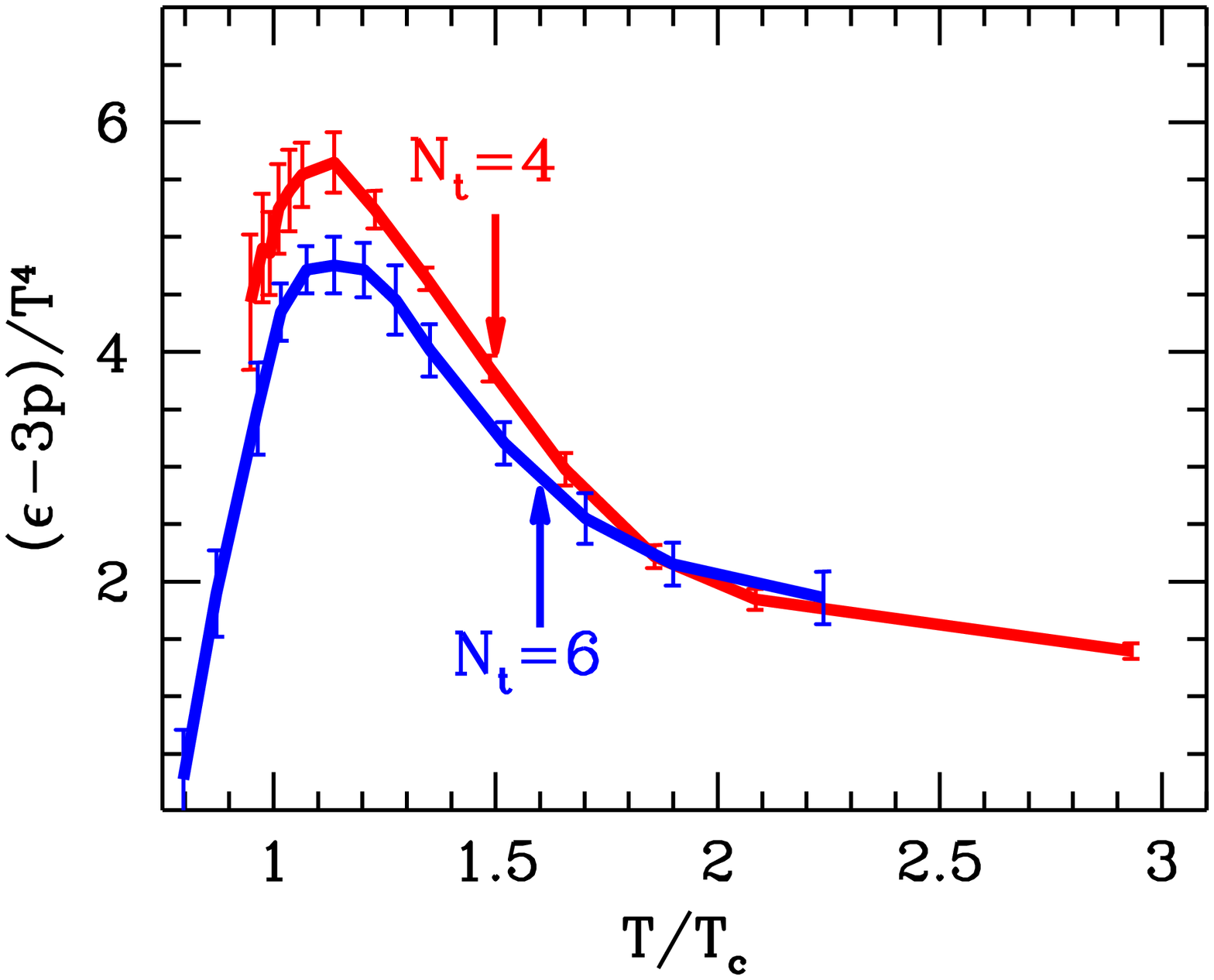,width=7.2cm}
\hspace{2cm}
\caption{\label{eos_e3p}
The interaction measure, the values are normalized by
$c_{cont}/c_{N_t}$ of the energy density. 
The labeling is the same as for Figure \ref{eos_pe}.
}
}
Let us present the results. In order to show how the different quantities
scale with the lattice spacing we show always $N_t$=4,6 results on the same
plot. In addition, in order to make the relationship with the continuum limit
more transparent we multiply the raw lattice results at finite temporal
extensions ($N_t$=4,6) with $c_{cont}/c_{N_t}$, where the c values are the
results in the free non-interacting plasma (Stefan-Boltzmann limit).  These c
values are summarized in Table \ref{ta:c_values} for the pressure, speed of
sound, and for the quark number susceptibility at $N_t$=4,6 and in the
continuum limit. By this multiplication  the lattice thermodynamic quantities
should approach the continuum Stefan-Boltzmann values for extreme large
temperatures.

Table \ref{p_values} contains our most important numerical results. We
tabulated the raw and normalized pressure values for both lattice
spacings and for all of our simulation 
points. This data set and eq. (\ref{eq:escs}) were used to obtain the following
figures.  Figure \ref{eos_pe} shows the equation of state on $N_t$=4,6 lattices.
The pressure (left panel) and $\epsilon$ (right panel) are presented as a
function of the temperature.  The Stefan-Boltzmann limit is also shown.  
On Figure \ref{eos_e3p} the interaction measure (i.e. $(\epsilon-3p)/T^4$) is
plotted, it is normalized by the corresponding $c_{cont}/c_{N_t}$ value of the
energy density.  
Figure \ref{eos_sc} shows the entropy density (left panel) and the speed of
sound (right panel), which can be obtained by using the pressure and energy
density data (c.f.  $sT$=$\epsilon$+$p$ and $c_s^2$=$dp$/$d\epsilon$) of the
previous Figure \ref{eos_pe}. Clearly, the uncertainties of the pressure and
those of the energy density cumulate in the speed of sound, therefore it is less
precisely determined.

\TABLE[t]{\label{p_values}
\begin{tabular}{|c|c|c|c||c|c|c|c|}
\hline
$\beta$    & $T/T_c$ & $p/T^4$ (raw)& $p/T^4$ (scaled) &
$\beta$    & $T/T_c$ & $p/T^4$ (raw)& $p/T^4$ (scaled) \\ \hline 
3.000 &$0.90$&$0.12(0.02)$&$0.07(0.01)$& 3.450 &  $0.80$&$0.07(0.11)$&$0.05(0.08)$ \\
3.150 &$0.95$&$0.32(0.07)$&$0.19(0.04)$& 3.500 &  $0.87$&$0.23(0.11)$&$0.15(0.08)$ \\
3.250 &$0.98$&$0.59(0.10)$&$0.34(0.06)$& 3.550 &  $0.96$&$0.59(0.12)$&$0.39(0.08)$ \\
3.275 &$0.99$&$0.73(0.10)$&$0.42(0.06)$& 3.575 &  $1.02$&$0.91(0.12)$&$0.60(0.08)$ \\
3.300 &$1.01$&$0.91(0.10)$&$0.52(0.06)$& 3.600 &  $1.07$&$1.29(0.13)$&$0.86(0.08)$ \\
3.325 &$1.04$&$1.13(0.10)$&$0.65(0.06)$& 3.625 &  $1.14$&$1.69(0.13)$&$1.12(0.09)$ \\
3.350 &$1.06$&$1.39(0.09)$&$0.79(0.05)$& 3.650 &  $1.20$&$2.10(0.14)$&$1.40(0.09)$ \\
3.400 &$1.14$&$2.04(0.10)$&$1.16(0.06)$& 3.675 &  $1.28$&$2.51(0.14)$&$1.66(0.10)$ \\
3.450 &$1.23$&$2.79(0.10)$&$1.59(0.06)$& 3.700 &  $1.35$&$2.88(0.15)$&$1.91(0.10)$ \\
3.500 &$1.34$&$3.56(0.11)$&$2.04(0.07)$& 3.750 &  $1.52$&$3.50(0.15)$&$2.32(0.10)$ \\
3.550 &$1.49$&$4.32(0.12)$&$2.47(0.07)$& 3.800 &  $1.70$&$3.99(0.16)$&$2.65(0.11)$ \\
3.600 &$1.66$&$4.96(0.12)$&$2.83(0.07)$& 3.850 &  $1.90$&$4.36(0.16)$&$2.89(0.11)$ \\
3.650 &$1.86$&$5.46(0.12)$&$3.12(0.07)$& 3.930 &  $2.24$&$4.82(0.17)$&$3.19(0.11)$ \\
3.700 &$2.09$&$5.84(0.12)$&$3.34(0.07)$& 4.000 &  $2.55$&$5.14(0.17)$&$3.41(0.11)$ \\
3.850 &$2.93$&$6.57(0.15)$&$3.75(0.09)$&       & 	&            &  \\
4.000 &$3.93$&$6.97(0.16)$&$3.98(0.09)$&       & 	&            &  \\
\hline                                     
\end{tabular}                              
\caption{                                  
Numerical values of the pressure for all of our simulation points.
The left column shows the $N_t$=4, whereas the right column shows
the $N_t$=6 data. Both the raw values and the ones scaled by 
$c_{cont}/c_{N_t}$ are given.
}
}
\FIGURE{
\epsfig{file=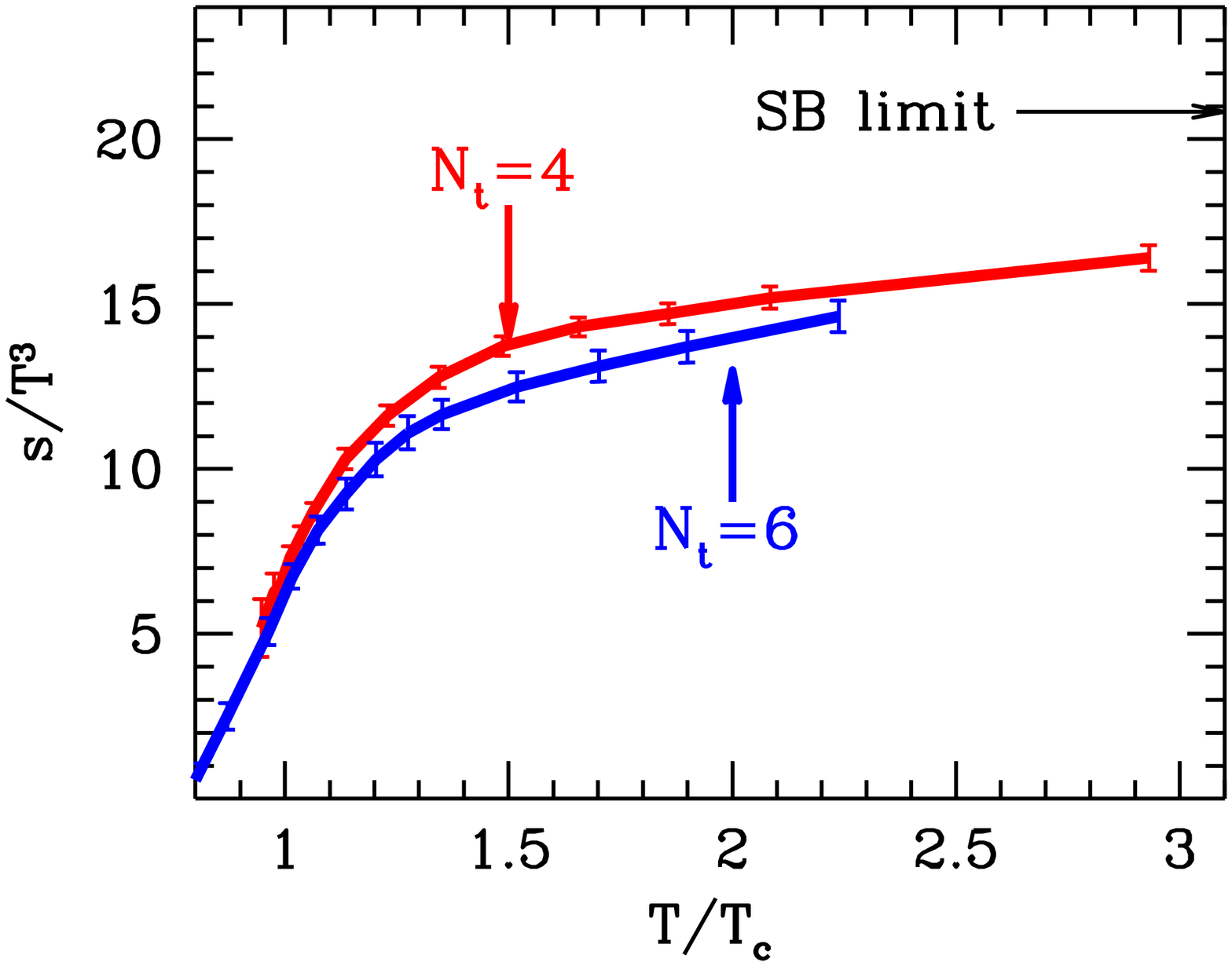,width=7.2cm}
\epsfig{file=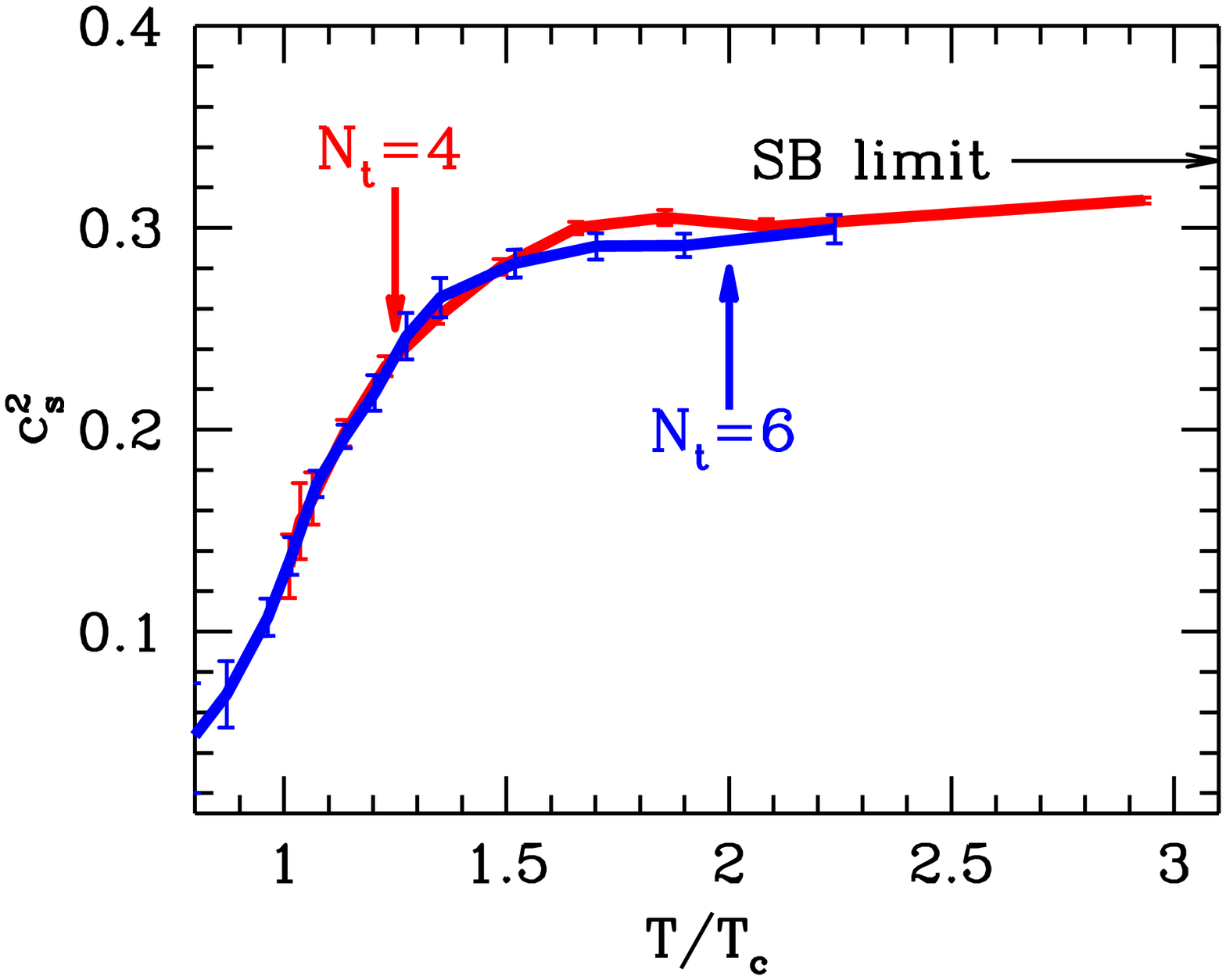,width=7.2cm}
\caption{\label{eos_sc}
The 
entropy density (left panel, normalized by $T^3$) and the
speed of sound (right panel). The labeling is the same as for Figure \ref{eos_pe}.
}
}

\FIGURE{
\epsfig{file=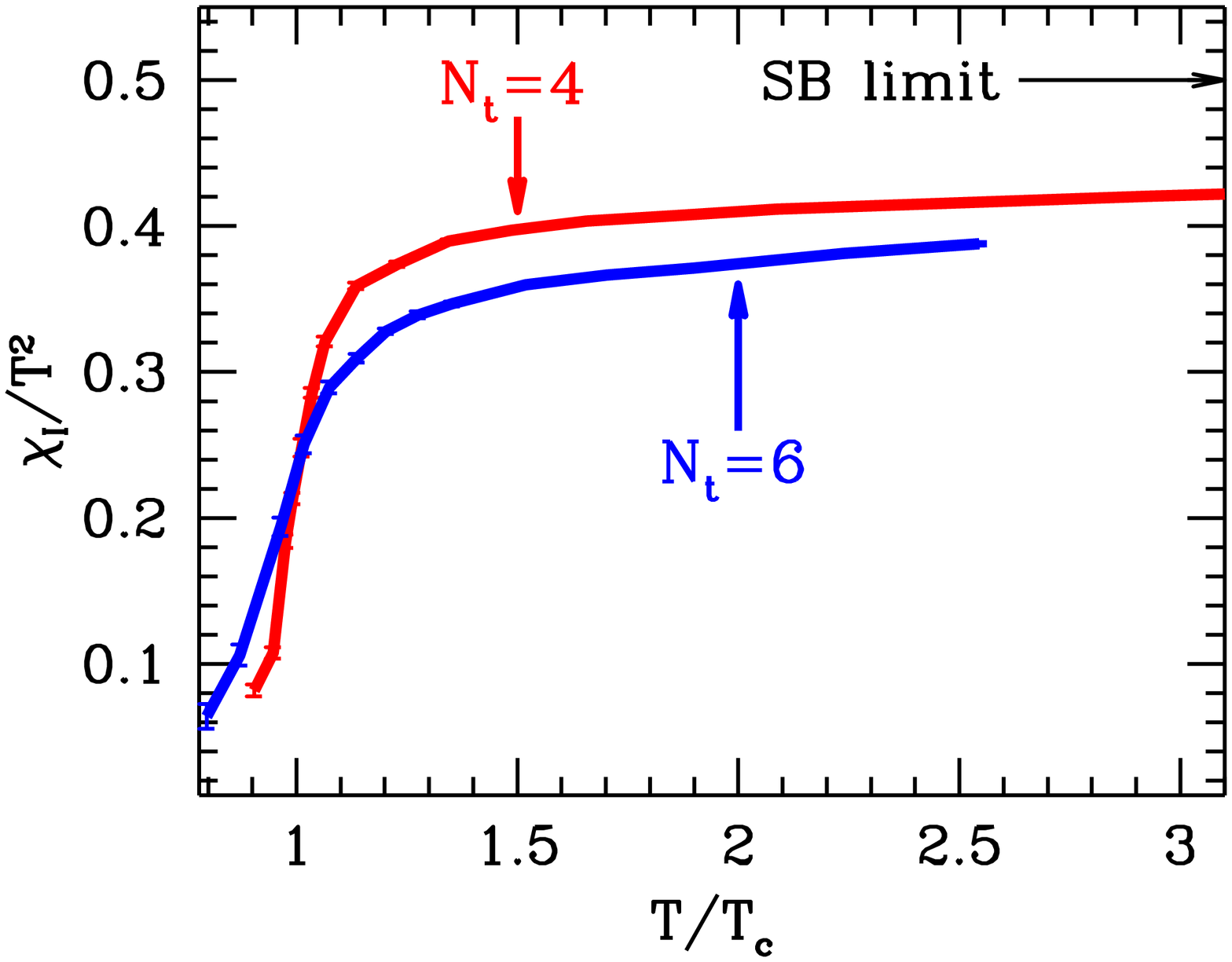,width=7.2cm}
\epsfig{file=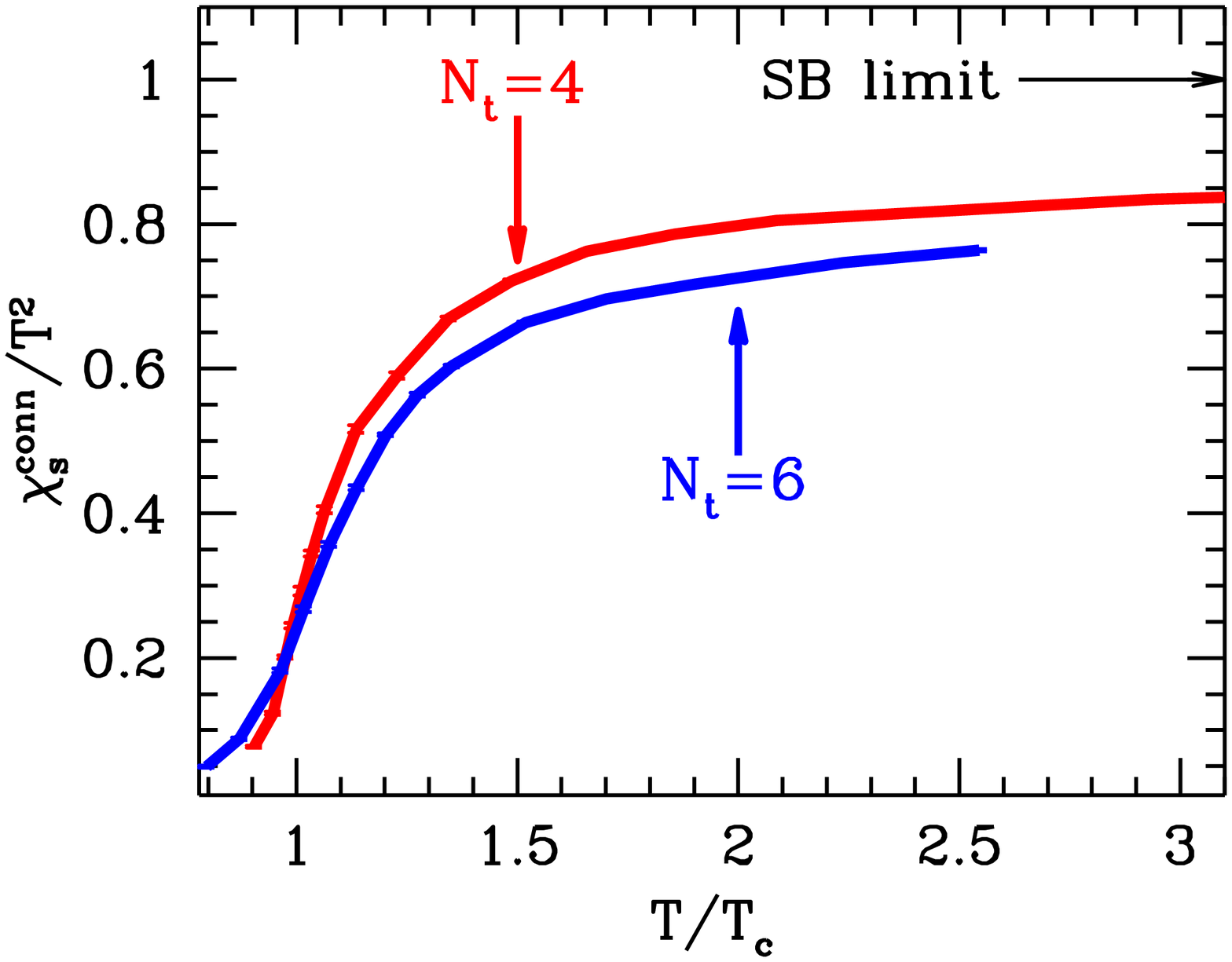,width=7.2cm}
\caption{\label{chi}
The isospin susceptibility 
(left panel, normalized by $T^2$) and the connected part of the strangeness susceptibility
(right panel, normalized by $T^2$). The labeling is the same as for Figure \ref{eos_pe}.
}
}
Light and strange quark number susceptibilities ($\chi_{ud}$ and $\chi_s$) are defined via
\cite{Bernard:2004je}
\begin{align}
\frac{\chi_{q}}{T^2}=\frac{N_t}{N_s^3}\left.\frac{\partial^2 \log Z}{\partial \mu_{q} ^2
}\right|_{\mu_{q}=0}, 
\end{align}
where $\mu_{ud}$ and $\mu_s$ are the light and strange quark chemical
potentials (in lattice units). With the help of the quark number operators 
\[Q_{q}=\frac{1}{4}\frac{\partial }{\partial
\mu_q}
\log \det (\Dsl+m_{q}),\] 
the susceptibilities can be written as
\[
\frac{\chi_{q}}{T^2}=\frac{N_t}{N_s^3}\left(\langle Q_q^2
\rangle_{\mu_q=0} + \left\langle \frac{\partial
Q_q}{\partial \mu_q}\right\rangle_{\mu_q=0}\right).
\]
The first term is usually referred as disconnected, the second as connected part. 
The connected part of the light quark number susceptibility is 2 times
the susceptibility of the isospin number ($\chi_I$). 
It is presented on the left panel of Figure \ref{chi}. 
For our statistics and evaluation method the disconnected 
parts are all consistent with zero and their value is far smaller than those
of the connected parts. 
The right panel of Figure \ref{chi} contains the connected part of the strange number
susceptibility.

\section{Summary, conclusion}

In this letter we presented results of a large scale numerical lattice study
on the thermodynamics of QCD. 

We determined the equation of state. Our analysis attempted to improve on
existing analyses by several means.  We used for the lightest hadronic degree
of freedom the physical pion mass. We used finer lattices with two different
sets of lattice spacings ($N_t$=4,6).  We kept our system on the line of
constant physics (LCP) instead of changing the physics with the temperature.
Due to our smaller lattice spacing and particularly due to our stout-link
improved fermionic action the unphysical pion mass splitting was much smaller
than in any previous analysis. We used an exact calculation algorithm instead
of an approximate one. Our scale was determined by a theoretically sound
quantity and not based on the string tension. 

We presented results for the pressure, energy density, entropy density, speed
of sound and on the isospin and strangeness susceptibilities.

Since the finite lattice spacing effects are quite different for different
temperatures a reliable continuum estimate can only be given if results on an
even finer lattice ($N_t$=8) were obtained.  This sort of analysis would be a
major step towards the final results for the equation of state.

{\bf Acknowledgments} 
We thank Claude Bernard and Carleton DeTar for providing the
pion splitting value from the MILC Asqtad result.
This research was partially supported by
OTKA Hungarian Science Grants No.\ T34980, T37615, M37071, T032501,
AT049652, and by DFG German Research Grant No.\ FO 502/1-1.
This research is part of the EU Integrated Infrastructure Initiative
Hadronphysics project under contract number RII3-CT-20040506078.
The computations were carried out at E\"otv\"os University
on the 330 processor PC cluster of the Institute for Theoretical Physics
and the 1024 processor PC cluster of Wuppertal University, using
a modified version of the publicly available MILC code \cite{MilcCode}
and a next-neighbour 
communication architecture
\cite{Fodor:2002zi}.

\bibliography{eos_mpi}

\end{document}